\documentstyle[preprint,aps,prc]{revtex}

\begin{document}
 
\newcommand{\beq}{\begin{equation}} \newcommand{\eeq}{\end{equation}}
\newcommand{\q}{{\bf q}} \newcommand{\h}{{\bf h}} \newcommand{\p}{{\bf
p}} \newcommand{\kk}{{\bf k}} \newcommand{\om}{{\omega}}
\newcommand{\RL}{{R_L(q,\omega)}} \newcommand{\RT}{{R_T(q,\omega)}}
\newcommand{\RTPH}{{R_T^{1p1h}(q,\omega)}}
\newcommand{\RTIA}{{R_T^{IA}(q,\omega)}}
\newcommand{\RTIN}{{R_T^{Int}(q,\omega)}}
\newcommand{\RTTB}{{R_T^{MEC}(q,\omega)}} \newcommand{\si}{{\bf
\sigma}} \newcommand{\ta}{{\bf \tau}} \newcommand{\JT}{{\bf j}({\bf
q})} \newcommand{\JC}{{\bf j}_C^{(1)}({\bf q})} \newcommand{\JM}{{\bf
j}_M^{(1)}({\bf q})} \newcommand{\JOB}{{\bf j}^{(1)}({\bf q})}
\newcommand{\JMEC}{{\bf j}^{(2)}({\bf q})} \newcommand{\JCON}{{\bf
j}_{Cont}^{(2)}({\bf q})} \newcommand{\JPI}{{\bf j}_{\pi}^{(2)}({\bf
q})} \newcommand{\JDE}{{\bf j}_{\Delta}^{(2)}({\bf q})}
\newcommand{\JPSC}{{\bf j}_{PS,C}^{(2)}({\bf q})}
\newcommand{\JPSP}{{\bf j}_{PS,\pi}^{(2)}({\bf q})}
\newcommand{\JPS}{{\bf j}_{PS}^{(2)}({\bf q})} \newcommand{\JV}{{\bf
j}_{V}^{(2)}({\bf q})} \newcommand{\JVS}{{\bf j}_{VS}^{(2)}({\bf q})}
\newcommand{\JMD}{{\bf j}_{MD}^{(2)}({\bf q})}
\newcommand{\EXP}{e^{\imath\q\cdot {\bf r}_i}} \draft

\title {The inclusive transverse response of nuclear matter}
\author{Adelchi Fabrocini} \address{ Department of Physics, University
of Pisa and Istituto Nazionale di Fisica Nucleare, Sezione di Pisa,
I-56100 Pisa, Italy}

 
\maketitle
 
\begin{abstract}
The electromagnetic inclusive transverse response of nuclear matter at
saturation density is studied within the correlated basis function
perturbation theory for momentum transfers $q$ from $300$ to
$550~MeV/c$.  The correlation operator includes a Jastrow component,
accounting for the short range repulsion, as well as longer range
spin, tensor and isospin ones. Up to correlated one particle-one hole
intermediate states are considered. The spreading due to the decay of
particle (hole) states into two particle-one hole (two hole-one
particle) states is considered via a realistic optical potential
model.  The Schiavilla-Pandharipande-Riska model for the two-body
electromagnetic currents, constructed so as to satisfy the continuity
equation with realistic $v_{14}$ potentials, is adopted.  Currents due
to intermediate $\Delta$-isobar excitations, are also included.
The global contribution of the two-body currents turns out to be
positive and provides an enhancement of the one-body transverse
response ranging from $\sim 20\%$ for the lower momenta to $\sim 10\%$
for the higher ones.  This finding is in agreement with the Green's
Function Monte Carlo studies of the transverse Euclidean response in
$A=3,4$ nuclei and contradicts previous results obtained within the
Fermi gas and shell models.  The tensor-isospin component of the
correlation is found to be the leading responsible for such a
behavior. The nuclear matter response is compared to recent
experimental data on $^{40}$Ca and $^{56}$Fe.
\end{abstract}
 


\narrowtext

{\bf I. INTRODUCTION}

 The cross section for inclusive electron scattering at intermediate
three-momentum transfers ($q\leq 600~MeV/c$) has been object of many
theoretical and experimental investigations. In the one-photon
exchange approximation the differential cross section is given by

\begin{equation}
\frac {d^2\sigma}{d\Omega d\om}=\sigma_M \left \{ \frac
{q^4_\mu}{q^4}R_L(q,\om)+[tan^2(\frac {\theta}{2})- \frac
{q^2_\mu}{2q^2}]R_T(q,\om)\right \}~~,
\label{eq:sigma}
\end {equation}

where $\sigma_M$ is the Mott cross section, $q^2_\mu=\om^2-q^2$ is the
squared four-momentum transfer, $\theta$ is the scattering angle and
$R_{L(T)}(q,\om)$ is the longitudinal (transverse) separated response.

 The total cross section is indeed well described by a simple Fermi
gas (FG) model\cite{Moniz71}, but the agreement disappears when the
longitudinal-transverse (L/T) separation
\cite{Barreau83,Deady83,Meziani84,Altemus80,Yates93,Zghiche94,Jourdan95,Jourdan96}
is carried out for medium-heavy nuclei.  $R_L$ is largely
overestimated by the FG model. However, the quenching of the
longitudinal response is now well understood in terms of short range
dynamical correlations, induced by the strong nucleon-nucleon (NN)
interaction, and of nucleon degrees of freedom alone.

 In Ref.\cite{R_L} a realistic model of correlated nuclear matter (NM)
was used to study $\RL$ at the NM empirical saturation density, in the
framework of the correlated basis function (CBF) theory. The density
dependent nuclear matter CBF results have then been used in
Ref.\cite{Jourdan96} to estimate the longitudinal response in
$^{12}$C, $^{40}$Ca and $^{56}$Fe in local density approximation
(LDA). The overall agreement with the experimental data was shown to
be satisfactory.

 The present understanding of $\RT$ is more uncertain. Recent
 experimental L/T separations in $^{40}$Ca \cite{Yates93,Jourdan96}
 have provided a transverse response lower than previous estimates
 \cite{Meziani84} at the quasielastic (QE) peak. On the other side,
 theoretical realistic calculations (feasible in the longitudinal
 case) have been so far prevented by the complicated structure of the
 transition operator, containing one- and two-body currents.  In light
 nuclei (A=3,4) only the Euclidean transverse response, with the full
 current operator, has been computed\cite{Euclid} using the exact
 Green's Function Monte Carlo (GFMC) technique and the realistic
 Argonne $v_{14}$ NN interaction\cite{Argonne14}.  GFMC cannot be
 presently adopted in heavier nuclei, so studies of $\RT$ in these
 systems either have considered only the easier to address one-body
 piece\cite{Depace93,Alberico89}, or have treated the two-body meson
 exchange currents (MEC) within independent particle models
 (IPM)\cite{VanOrden81,Alberico84,Amaro94}.  The MEC were found to
 substantially increase the one-body response in Ref.\cite{Euclid},
 whereas the IPM calculations of Ref.\cite{Amaro94} point to a slight
 reduction. It is worth noticing that the aforementioned latest heavy
 nuclei separated $\RT$ show a good agreement with theoretical
 responses containing only one-body currents\cite{Jourdan96},
 downplaying the role of the MEC and in contrast with both the light
 nuclei case and the old $^{40}$Ca data.

 Aim of this work is to use CBF theory to compute the symmetric
 nuclear matter transverse response in order to ascertain how it is
 affected by the NN correlations. Particular attention will be devoted
 to their influence on the MEC contribution. The results presented in
 the paper have been obtained within the exchange current operator
 model developed by Schiavilla, Pandharipande and Riska (SPR) in
 Ref.\cite{SPR_MEC}.  The SPR model satisfies the continuity equation
 linking the current to the NN interaction and contains intermediate
 $\Delta$-isobar excitations.  For the sake of comparison, also the
 standard one-pion exchange currents have been used.

 In nuclear matter, CBF calculations are based upon a set of {\it
 correlated} wave functions

\begin{equation}
|n\rangle={\cal S}\left [ \prod_{i<j} f(i,j)\right ] |n\rangle_{FG}
~~,
\label{eq:|n>}
\end {equation}

obtained by applying a symmetrized product of two-body correlation
operators, $f(i,j)$, to the FG states $|n\rangle_{FG}$. An effective
structure for $f(i,j)$ is

\begin{equation}
f(i,j)=\sum_{q=1,6}f^{(q)}(r_{ij})O^{(q)}_{ij}\: ,
 \:\,O^{(q=1,6)}_{ij}=(1,\si_i\cdot\si_j,S_{ij})\otimes(1,\ta_i\cdot\ta_j)
 ~~,
\label{eq:f_ij}
\end {equation}
 
 $S_{ij}$ being the tensor operator; this operatorial dependence
 resembles the structure of the NN interaction. Additional spin-orbit
 correlations are often considered in realistic NM groud state
 studies, however, they will not be taken into account in this work.
 $f(i,j)$ depends upon a set of variational parameters, which are
 fixed by minimizing the expectation value of a realistic, non
 relativistic hamiltonian on the correlated groud state. The
 g.s. energy is calculated via the Fermi hypernetted chain (FHNC)
 cluster summation technique\cite{RosatiFHNC} and the operatorial
 contributions ($q>1$) are implemented by the single operator chain
 (SOC) (and successive improvements)
 approximation\cite{WiringaSOC,Fiks88}.
 We have used the correlation corresponding to the Argonne
 $v_{14}$+Urbana VII three-nucleon interaction model of
 Ref.\cite{Fiks88}.  A model with a weaker tensor force (the Urbana
 $v_{14}$+TNI interactionl\cite{Urbana14}) was adopted for the CBF
 studies of $\RL$ in Ref.\cite{R_L} and of the L/T spin responses in
 Ref.\cite{R_spin}.

 The correlation given in Eq.(\ref{eq:f_ij}) contains a $q=1$, scalar
 (or Jastrow) component, $f^J(r)$, almost vanishing at short distances
 so that configurations where two nucleons are close enough are
 essentially suppressed in the wave function because of the short
 range NN repulsion. $f^J(r)$ heals to unity at large distances.  The
 most important among the remaining components are, by far, the $q=5$,
 spin-isospin, $f^{\sigma\tau}(r)$, and the $q=6$, tensor-isospin,
 $f^{t\tau}(r)$, ones, which are related to the one-pion exchange
 (OPE), long range part of the potential\cite{Fiks88}.

 The response is computed by considering up to correlated one
 particle-one hole ({$1p1h$) intermediate states. Admixtures of
 correlated particle (hole) states with $2p1h$($2h1p$) ones are
 accounted for by the optical potential model of
 Ref.\cite{Fantoni87}. This model has been proven to be fairly
 accurate in the momentum region scanned here, at least for the
 longitudinal response \cite{R_L}. No correlated $2p2h$ intermediate
 states have been considered, since they contribute mainly at larger
 energies than those of the QE peak\cite{Amaro94}, which is the focus
 of this work.  A peak due the excitation of $\Delta$-resonances is
 also present in the transverse response at high energies.  The
 $\Delta$ peak is however well distinct from the quasielastic one and
 it has not been studied in the paper. It would have required the
 introduction of $\Delta$-degrees of freedom in the nuclear wave
 function.

 The plan of the paper is as follows. The CBF approach to the
 transverse response is sketched and the current operators are briefly
 described in Section II .  Section III is devoted to the discussion
 of the matrix elements of the current. The numerical results are
 presented and discussed in Section IV, together with a comparison
 with some of the available experimental data.  Finally, some
 conclusions is drawn.

{\bf II. THE CURRENT OPERATOR AND THE CBF THEORY OF THE TRANSVERSE
RESPONSE}

 The transverse response $\RT$ is given by

\begin{equation}
\RT={1\over A}\sum_n \vert\langle 0|\JT | n\rangle\vert ^2 \delta (\om
- \om_n)~~, _\label{eq:TSF1}
\end{equation} 

where the sum goes over the intermediate excited states $| n\rangle$,
having excitation energy $\om_n$, and $\JT$ is the electromagnetic
current operator.  $\JT$ has contributions from one-body, $\JOB$, and
two-body exchange currents, $\JMEC$. In the impulse approximation (IA)
only $\JOB$ is retained.
 
$\JOB$ is the sum of the convection, $\JC$, and magnetization, $\JM$,
 terms.  However, it is known that the convection current contributes
 significantly to the non energy weighted sum of the response at very
 low $q$-values, where it becomes dominant \cite{Schiavilla89}.  At
 momentum transfers $\geq 150~MeV/c$ it provides only a few percent of
 the total sum, with a correction going as $1/q^2$. For this reason,
 $\JC$ has not been considered here and we have approximated

\begin{equation}
\JOB\sim\JM = \imath G(q,\om) \frac {\mu_0}{e} \sum_{i=1,A} \EXP \left
[ \mu_p \frac {1+\ta_{i,z}}{2}+\mu_n \frac {1-\ta_{i,z}}{2} \right ]
\si_i\times\q ~~, _\label{eq:J_OB}
\end{equation} 

 where $\mu_0$ is the nuclear magneton and $\mu_{p,n}$ are the nucleon
 magnetic moments. The dipole parametrizion has been used for the
 nucleon form factor $G(q,\om)$, with a scale parameter of 839 $MeV$.
 Studying the dependence of the response to the different available
 parametrizations of the electromagnetic nucleon form factors is
 beyond the scope of the present paper.

 $\JM$ can be written in terms of the isoscalar and isovector
 transverse spin fluctuation operators, $\rho_{T,\si}^{\ta=0,1}(\q)$,
 
\begin{equation}
\rho_{T,\si}^{0(1)}(\q) = \frac {1}{\sqrt 2} \sum_{i=1,A} \EXP (
 \si_i\times\q )\,( \ta_{i,z}) ~~, _\label{eq:rho_spin}
\end{equation} 

 and, correspondingly, the IA electromagnetic response is related to
 the spin structure functions, $S_{T,\si}^{0,1}(\q,\om)$, by

\begin{equation}
\RTIA = \left ( \frac {\mu_0}{e{\sqrt 2}} \right )^2 \vert G(q,\om
)|^2 \left [ \mu_+^2 S_{T,\si}^0(\q,\om)+\mu_-^2
S_{T,\si}^1(\q,\om)\right ] ~~, _\label{eq:ST_IA}
\end{equation} 

 where $\mu_\pm=\mu_p\pm\mu_n$. $S_{T,\si}^{0,1}(\q,\om)$ have been
 computed in Ref.\cite{R_spin} in FHNC/SOC. So, the details of the CBF
 calculation of the IA transverse response will not be discussed here.

 Most of the studies on exchange currents are based on simple meson
 exchange mechanims, in a non-relativistic description of the
 nucleon-meson-$\Delta$ systems\cite{RiskaMEC}. In this case MEC
 associated with one-pion exchange are given by the sum of three
 terms, obtained by the corresponding Feynman diagrams: the contact,
 or seagull term, $\JCON$, the pionic, or pion in flight term, $\JPI$,
 and the $\Delta$ term, $\JDE$, associated with the excitation of
 intermediate $\Delta$ resonances.  Analogous contributions arise from
 one-rho exchange. Their expressions can be found in several papers,
 and, among them, in Ref.\cite{SPR_MEC}. These currents are consistent
 with NN potentials based only on one-boson exchange (OBE), so the
 continuity equation,

\begin{equation}
\nabla\cdot{\bf j}({\bf x};{\bf r}_1,{\bf r}_2)+ \imath\left [v({\bf
r}_1,{\bf r}_2),\rho({\bf x})\right ]= 0 ~~, _\label{eq:cont_eq}
\end{equation} 

 with realistic potentials is not satisfied.

 In Ref.\cite{SPR_MEC}, the Authors have derived exchange currents
 satisfying the continuity equation with Argonne and Urbana $v_{14}$
 potentials. The current was separated in the sum of a model
 independent (MI) component, constructed from the NN interaction, and
 a model dependent (MD) one, purely transverse and given by the
 previously introduced $\pi$ and $\rho$ MEC with intermediate $\Delta$
 excitations.

 The most important MI currents are those associated with the isospin
 dependent parts of the NN interaction.  The corresponding current
 operators, $\JPS$, $\JV$ and $\JVS$, are given in
 Ref.\cite{SPR_MEC}. They are expressed in terms of the Fourier
 tranforms of the $v^\ta(r), v^{\si\ta}(r)$ and $v^{t\ta}(r)$ parts of
 the potential\cite{SPR_MEC} and reduce to the standard $\pi$ and
 $\rho$ MEC if the OBE interaction is used. Argonne $v_{14}$ has been
 used to generate the MI currents: they are close to those due to
 $\pi$ and $\rho$ exchanges, even if the interaction does not strictly
 have the OBE form.

 The MD part of the current was associated to $\pi$ and $\rho$ MEC
 with intermediate $\Delta$ excitations and to $\rho\pi\gamma$ and
 $\om\pi\gamma$ currents, all of them treated within the OBE
 model. The last two currents will not be considered here.  Later,
 $\Delta$s were explicitly included in the wave
 function\cite{Schiavilla92}, instead of introducing effective
 two-body operators acting on nucleon coordinates, as in standard
 first order perturbation theory.  These components were generated by
 transition correlation operators,
 $U^{TR}_{ij}=U^{N\Delta}_{ij}+U^{\Delta
 N}_{ij}+U^{\Delta\Delta}_{ij}$, acting on a realistic nuclear wave
 functions and obtained via the Argonne $v_{28}$\cite{Argonne14}
 potential, which contains $\Delta$ degrees of freedom. Hence,
 N$\rightarrow\Delta$ and $\Delta\rightarrow\Delta$ transition
 operators were introduced in the one-body current. The $\Delta$
 contribution to low-energy electroweak transitions was found to be
 smaller than that of perturbative theories as a consequence of this
 more realistic approach. It was also noticed that a first order
 perturbative treatment, consistent with that of
 Ref.\cite{Schiavilla92}, is obtained by using
 
\begin{equation}
u_{PT}^{\si (t)\ta II}(r_{ij})=\frac {v^{\si(t)\ta
II}(r_{ij})}{m_N-m_{\Delta}} ~~, _\label{eq:u_PTst}
\end{equation} 

 for the N$\rightarrow\Delta$ spin- and tensor-isospin transition
 correlations, where $v^{\si (t)\ta II}$ are the corresponding
 transition components of the Argone $v_{28}$ potential.

 The intermediate $\Delta$ current generated in this way will be used.
 Its configuration space expression\cite{Schiavilla_priv} is

\begin{eqnarray}
\nonumber \JMD & = & \imath \frac {2}{9} {G_\Delta(q,\om)} \mu_{\gamma
N \Delta} \sum_{i<j} [ \EXP \{ 4\ta_{j,z} [ ( u_{PT}^{\si\ta II}(r) -
u_{PT}^{t\ta II}(r) ) \si_j\times \q \\ \nonumber & + & 3 u_{PT}^{t\ta
II}(r) \hat{\bf r}\times\q (\si_j\cdot \hat{\bf r}) ] - (\ta_i \times
\ta_j )_z [ ( u_{PT}^{\si\ta II}(r) - u_{PT}^{t\ta II}(r) )
(\si_i\times\si_j)\times \q \\ & + & 3 u_{PT}^{t\ta II}(r)
(\si_i\times\hat{\bf r})\times\q (\si_j\cdot\hat{\bf r}) ] \} + i
\rightleftharpoons j ] ~~.
\end{eqnarray}

 ${\bf r}={\bf r}_{ij}$ and $\mu_{\gamma N \Delta}=3\mu_0$ is the
 transition magnetic moment adopted in Ref.\cite{Schiavilla92}, whose
 value is $\sim 30\%$ smaller than the static quark model
 prediction\cite{Brown75}. For the $\Delta$ form factor we take

\begin{equation}
 G_\Delta(q,\om) = \left ( 1 + \frac {q_\mu^2}{\Lambda_1^2} \right
 )^{-2} \left ( 1 + \frac {q_\mu^2}{\Lambda_2^2} \right )^{-1/2} ~~,
 _\label{eq:G_D}
\end{equation} 

 where the cutoff masses are $\Lambda_1=1196~MeV$ and
 $\Lambda_2=843~MeV$.

 As a result of the separation of the current into one- and two-body
 pieces, the response is given by

\begin{equation}
\RT= \RTIA+\RTIN+\RTTB ~~,
\label{eq:TSF2}
\end {equation}

 where $\RTIN$ is the interference contribution between $\JOB$ and
 $\JMEC$,

\begin{equation}
\RTIN={1\over A}\sum_n [\langle 0|\JOB | n\rangle\langle
n|(\JMEC)^\dagger | 0\rangle+c.c.]  \delta (\om - \om_n)~~,
_\label{eq:TSF3}
\end{equation} 

 and

\begin{equation}
\RTTB={1\over A}\sum_n \vert\langle 0|\JMEC | n\rangle |^2 \delta (\om
- \om_n)~~.  _\label{eq:TSF4}
\end{equation} 

 $\RTIN$ and the leading term in $\RTTB$ only have been computed.

 Let's now shortly discuss the most important aspects of the CBF
 perturbative expansion. The goal of using a correlated basis,
 embodying directly into the states some of the relevant physical
 effetcs (as the short range repulsion), is to obtain a rapidly
 converging expansion. The obvious price to be paid lies in the
 greater difficulty in evaluating matrix elements, even at the zeroth
 order. However, nuclear matter studies demonstrated that CBF based
 perturbation theory is actually fast converging, in the sense {\it
 i}) that for many observables the zeroth order is already a good
 estimate and {\it ii}) that the inclusion of the first CBF
 intermediate states is often sufficient to provide a quantitative
 agreement with the empirical values. This happens for ground state
 properties, as the energy\cite{Fabrocini93} and the momentum
 distribution\cite{Fantoni84}, as well as for the already mentioned
 longitudinal and spin responses and the nucleon spectral
 function\cite{Benhar89}. The inclusive electron scattering cross
 section at high momentum transfers has been computed in
 Ref.\cite{Benhar91}, using the CBF spectral function, and
 satisfactorily compared with that extrapolated from data on
 laboratory nuclei.

 Hence, relying on these facts, we will compute $\RT$ by inserting in
 the intermediate states summation of Eq.(\ref{eq:TSF1}) only
 correlated $1p1h$ states

\begin{equation}
|\p\h\rangle={\cal S}\left [ \prod_{i<j} f(i,j)\right ]
|\p\h\rangle_{FG} ~~.
\label{eq:|ph>}
\end {equation}
 
 This choice is also justified on the ground that we are interested
 mainly in the QE peak. The quantitative study of the large energy
 region would have required the insertion of higher excited states
 with both nucleonic (as $2p2h$) and explicit $\Delta$ isobar
 excitations (as $\Delta$-$h$).

 The $1p1h$ transverse response is then given by

\begin{equation}
\RTPH={1\over A}\sum_{ph} \vert\langle 0|\JT |\p\h \rangle\vert ^2
\delta (\om - e_p+e_h)~~, _\label{eq:TSFPH}
\end {equation}
 
 where $e_{x=p,h}=\langle {\bf x} | H | {\bf x} \rangle - \langle 0 |
 H | 0 \rangle $ is the CBF variational (or zeroth order) single
 particle energy of the $x$-state.

 The $1p1h$ response has sharp energy boundaries, qualitatively like
 the Fermi gas, ruled by the real part of the CBF optical potential,
 $U_k=e_k-\hbar^2 k^2/2m_N$\cite{Fabrocini93}.  The decay of $1p1h$
 states into $2p2h$ ones has the main effect of introducing a large
 energy tail and redistributing the strength. In the perturbative
 expansion this is accounted for by self-energy insertions on top of
 the particle or hole line.  The real part of the self-energy provides
 a perturbative correction, $\delta e_x$, to the variational single
 particle energy and the imaginary part induces the spreading of the
 response to high $\om$-values. The microscopically computed CBF
 self-energy was used in the longitudinal response calculation of
 Ref.\cite{R_L}. In Ref.\cite{Fantoni87} these corrections were
 estimated by folding the $1p1h$ response with a width $W(\om)$ given
 by the imaginary part $W_0(\om)$ of the optical potential divided by
 the nucleon effective mass,

\begin{equation}
\RT=\frac {1}{\pi}\int d\om'R_T^{1p1h}(q,\om') \frac
{W(\om)}{(\om-\om')^2+W(\om)^2}~~, _\label{eq:FOLD}
\end {equation}

 with $W_0(\om)\sim 11\om^2/(4900+\om^2)$, in MeV. This procedure,
 numerically much less involved and equivalent to retain the on-energy
 shell part of the self-energy only, was checked to be reliable in the
 momentum region of interest\cite{R_L} and, for this reason, has been
 adopted here.

{\bf III. CBF MATRIX ELEMENTS OF THE CURRENT}

 This Section will focus on the main features of the MEC CBF matrix
 elements.  As far as the MI currents is concerned, only the leading
 $\JPS$ will be discussed. However, the contributions of the less
 influent $\JV$ and $\JVS$ have been evaluated and will be presented
 in the next Section.

 The configuration space PS current is written as a sum of two terms,
 $\JPSC$ and $\JPSP$. The first one coincides with $\JCON$ for the
 one-pion exchange potential, the latter with $\JPI$. Their
 expressions are
 
\begin{equation}
\JPSC = C_{PS}\sum_{i<j} 3 (\ta_i \times \ta_j )_z \{ \EXP g_{PS}(r)
 \si_i (\si_j\cdot \hat{\bf r}) + i \rightleftharpoons j \}~~,
 _\label{eq:JPS_1}
\end {equation}

 and

\begin{eqnarray}
\nonumber \JPSP & = & C_{PS}\sum_{i<j} 3 (\ta_i \times \ta_j )_z
e^{\imath \q \cdot {\bf R}} \{ \frac {G_{PS,1}({\bf r})}{r^2} [
\si_i(\si_j\cdot\hat{\bf r}) + \si_j(\si_i\cdot\hat{\bf r}) + \hat{\bf
r}(\si_i\cdot\si_j) ] \\ \nonumber & + & \imath \frac {G_{PS,2}({\bf
r})}{r} \si_i(\si_j\cdot\q) - \imath \frac {G_{PS,3}({\bf r})}{r}
\si_j(\si_i\cdot\q) \\ \nonumber & - & \imath \frac {G_{PS,4}({\bf
r})}{r} \hat{\bf r} (\si_i\cdot \hat{\bf r})(\si_j\cdot\q) + \imath
\frac {G_{PS,5}({\bf r})}{r} \hat{\bf r} (\si_j\cdot \hat{\bf
r})(\si_i\cdot\q) + \\ & - & G_{PS,6}({\bf r}) \hat{\bf r}(\si_i\cdot
\q)(\si_j\cdot\q) - \imath \frac {G_{PS,7}({\bf r})}{r^2} \hat{\bf r}
(\si_i\cdot \hat{\bf r})(\si_j\cdot\hat{\bf r}) \}~~.
\end{eqnarray}

 The functions $g_{PS}(r)$ and $G_{PS,\alpha=1,7}({\bf r})$ are
 defined in Ref.\cite{SPR_MEC}, ${\bf R}=({\bf r}_i+{\bf r}_j)/2$ and
 $C_{PS}=G(q,\om)$.

 In CBF the non diagonal matrix elements, ${\bf \xi}_{PS,C(\pi)}({\bf
  q};{\bf p},{\bf h})= \langle 0 | {\bf j}_{PS,C(\pi)}^{(2)}({\bf q})
  | {\bf p},{\bf h} \rangle$ are computed by cluster expansions in
  Mayer like diagrams, built up by dynamical correlations,
  $f^{(q)}f^{(p)}-\delta_{q1}\delta_{p1}$, and by exchange
  links. Infinite classes of cluster terms containing Jastrow
  correlations are summed by FHNC. Less massive summations, similar to
  SOC, can be performed for the operatorial
  correlations\cite{R_L,Fantoni87}.

 In the case of simple Jastrow correlated wave functions
  ($f^{q>1}=0$), ${\bf \xi}^J_{PS,C}$ is given by

\begin{eqnarray}
\nonumber
{\bf \xi}_{PS,C}^J({\bf q};{\bf p},{\bf h}) & = & \delta_{{\bf q}-{\bf
 p}+{\bf h}} \imath C_{PS} \frac { 6 \ta_z}{\sqrt{D(p)D(h)}} \\ & &
 \rho_0 \int d{\bf r} g_{PS}(r)g_{cc}(r) \left \{ e^{\imath \p \cdot
 {\bf r}} (\hat{\bf r} - \imath \si \times \hat{\bf r}) + e^{\imath \h
 \cdot {\bf r}} (\hat{\bf r} + \imath \si \times \hat{\bf r}) \right
 \}~~,
\label{eq:csiJ_C}
\end{eqnarray}

 where $\ta_z=\ta_{p,z}=\ta_{h,z}$, $\si=\si_{p}=\si_{h}$,
 $D(x=p,h)=1-{X}_{cc}(x)$ (see Ref.\cite{Fantoni87} for the definition
 of ${X}_{cc}$), $g_{cc}(r)$ is the exchange FHNC partial radial
 distribution function\cite{RosatiFHNC} and $\rho_0$ is the nuclear
 matter density.  At the lowest order of the cluster expansion,
 $D(x)=1$ and $g_{CC}(r)=-[f^J(r)]^2 l(k_Fr)/4$, where $k_F$ is the
 Fermi momentum and $l(x)=3[sin(x)-xcos(x)]/x^3$ is the Slater
 exchange function. So, Jastrow correlations suppress the short range
 part of the pion-like exchange interaction. This effect is visualized
 in Fig.\ref{fig:1}, where the Argonne $v_{14}$ $g_{PS}$ and the FHNC
 $g_{cc}$ functions are shown. Moreover, the Figure compares
 $g_{cc}(FHNC)$ with its uncorrelated Fermi gas counterpart,
 $g_{cc}(FG)=-l(k_Fr)/4$, to stress their different short distance
 behaviors.  Finally, the simple $\pi$ exchange $g_{PS}(\pi)
 (f^2_\pi/4\pi=0.081)$ is given.  As it was already stated, it results
 to be close to the full $g_{PS}$.

  Eq.(\ref{eq:csiJ_C}) sums all cluster diagrams factorizable in
 products of {\it dressed} two--body diagrams.  Non factorizable
 diagrams involving three particles have also been taken into account,
 even if they do not appear in the expression.
  
 Operatorial correlations were introduced in Ref.\cite{R_spin} for the
 spin responses within both the dressed two-body (D2B) and the SOC
 approximations.  In the first approximation cluster diagrams
 containing Jastrow correlations are summed to all orders, while
 contributions from the other components are evaluated at the two-body
 level (it corresponds to the $W_0$ approximation of
 Ref.\cite{WiringaSOC}). D2B came out to be very accurate and it has
 been used here to evaluate the IA response.  A linearized version of
 this approximation (D2B/L), including only contributions linear in
 $f^{\sigma\tau,t\tau}$, has been adopted for the MEC matrix
 elements. However, in some cases quadratic terms, as well as those
 containing the other operatorial components, have been computed and
 checked to give small corrections.

 Following the above scheme for the treatment of the $q>1$
 correlations, ${\bf \xi}_{PS,C}$ has an exchange part, ${\bf
 \xi}^{exch}_{PS,C}$, obtained by the replacements
 
\begin{equation}
g_{cc}(r)\hat{\bf r} \rightarrow g_{cc}(r)\hat{\bf r} \left \{ 1+ 2
 \frac {f^{\sigma\tau}(r)-4f^{t\tau}(r)}{f^J(r)}\right \} ~~,
\label{eq:csi_C1}
\end{equation}

\begin{equation}
g_{cc}(r)\imath(\si\times\hat{\bf r}) \rightarrow
g_{cc}(r)\imath(\si\times\hat{\bf r}) \left \{ 1+ 2 \frac
{f^{\sigma\tau}(r)-f^{t\tau}(r)}{f^J(r)}\right \} ~~,
\label{eq:csi_C2}
\end{equation}

 in Eq.(\ref{eq:csiJ_C}), and a direct one

\begin{eqnarray}
\nonumber {\bf \xi}^{dir}_{PS,C}({\bf q};{\bf p},{\bf h}) & = &
\delta_{{\bf q}-{\bf p}+{\bf h}} \imath C_{PS} \frac { 6
\ta_z}{\sqrt{D(p)D(h)}} \\ & & \rho_0 \int d{\bf r} g_{PS}(r)[
g_{dd}(r) + g_{de}(r) ] e^{\imath \q \cdot {\bf r}} 2\imath (\si
\times \hat{\bf r}) \left \{ \frac
{f^{\sigma\tau}(r)-f^{t\tau}(r)}{f^J(r)} \right \} ~~,
\end{eqnarray}

 where $g_{dd,de}(r)$ are the direct-direct and direct-exchange FHNC
 partial distribution functions.

 $\JPSC$ is the leading two-body current and it is the only one used
 to estimate $\RTTB$ of Eq.\ref{eq:TSF4}.

 $\JPSP$ is the sum of several components. Following an obvious
 notation, we will refer to the corresponding parts of its matrix
 element as ${\bf \xi}_{PS,\pi}(G_{PS,\alpha=1,7})$. They are given in
 the Appendix.

 In presenting the matrix elements of the $\JMD$ current, ${\bf
 \xi}_{MD}({\bf q};{\bf p},{\bf h})= \langle 0 | \JMD | {\bf p},{\bf
 h} \rangle$, we introduce, for the sake of brevity, the notations

\begin{equation}
 C_\Delta=\frac {2}{9} {G_\Delta(q,\om)} \mu_{\gamma N \Delta} ~~,~~
 f_1^\Delta (r)= u_{PT}^{\si\ta II}(r) - u_{PT}^{t\ta II}(r) ~~,~~
 f_2^\Delta (r)= 3 u_{PT}^{\si\ta II}(r) ~~.
\label{eq:C_delta}
\end{equation}

 ${\bf \xi}_{MD}$ is given by

\begin{equation}
{\bf \xi}_{MD}({\bf q};{\bf p},{\bf h}) = {\bf \xi}_{MD}^{dir}({\bf
q};{\bf p},{\bf h}) + {\bf \xi}_{MD}^{exch}({\bf q};{\bf p},{\bf h})
\label{eq:csi_MD}
\end{equation}
 
  where

\begin{eqnarray}
\nonumber {\bf \xi}_{MD}^{dir}({\bf q};{\bf p},{\bf h}) & = &
\delta_{{\bf q}-{\bf p}+{\bf h}} \imath C_\Delta q \frac { 4
\ta_z}{\sqrt{D(p)D(h)}} \rho_0 \int d{\bf r} [ g_{dd}(r) + g_{de}(r) ]
e^{\imath {\q} \cdot {\bf r}} \\ \nonumber & & \left \{ f^\Delta_1(r)
\left [ ( \si \times \hat{\q} ) \left ( 1 + \frac {2 f^{\si\ta}(r)
+f^{t\ta}(r)} {f^J(r)} \right ) - 3 (\si \cdot \hat{\bf r}) ( \hat{\bf
r} \times \hat{\q} ) \frac {f^{t\ta}(r)}{f^J(r)} \right ] \right.  \\
\nonumber & + & \left.  f^\Delta_2(r) \left [ ( \si \times \hat{\q} )
\left ( \frac {2 f^{\si\ta}(r) - f^{t\ta}(r)}{f^J(r)} \right ) + (\si
\cdot \hat{\bf r}) ( \hat{\bf r} \times \hat{\q} ) \left ( 1 + \frac {
f^{t\ta}(r)-f^{\si\ta}(r)}{f^J(r)} \right ) \right ] \right \} \\
\nonumber & + & \delta_{{\bf q}-{\bf p}+{\bf h}} \imath C_\Delta q
\frac { 4 \ta_z}{\sqrt{D(p)D(h)}} \rho_0 \int d{\bf r} [ g_{dd}(r) +
g_{de}(r) ] \left [ ( \si \times \hat{\bf q} ) + (\si \cdot \hat{\bf
r}) ( \hat{\bf r} \times \hat{\bf q} ) \right ] \\ & + & \left \{ 3
f^\Delta_1(r) \frac {f^{\si\ta}(r)}{f^J(r)} + f^\Delta_2(r) \left (
\frac {2 f^{t\ta}(r)+f^{\si\ta}(r)}{f^J(r)} \right ) \right \} ~~,
\label{eq:csi_MDd}
\end{eqnarray}

\begin{eqnarray}
\nonumber {\bf \xi}_{MD}^{exch} ({\bf q};{\bf p},{\bf h}) & = &
 \delta_{{\bf q}-{\bf p}+{\bf h}} \imath C_\Delta q \frac { 2
 \ta_z}{\sqrt{D(p)D(h)}} \\ \nonumber & & \rho_0 \int d{\bf r}
 g_{cc}(r) \left [ e^{\imath \p \cdot {\bf r}} + e^{\imath \h \cdot
 {\bf r}} \right ] \left \{ f^\Delta_1(r) \left [ ( \si \times
 \hat{\q} ) \left ( 4 + 3 \frac {8 f^{\si\ta}(r) -f^{t\ta}(r)}
 {f^J(r)} \right ) \right. \right.  \\ \nonumber & - & \left. 9 (\si
 \cdot \hat{\bf r}) ( \hat{\bf r} \times \hat{\q} ) \frac
 {f^{t\ta}(r)}{f^J(r)} \right ] + f^\Delta_2(r) \left [ ( \si \times
 \hat{\q} ) \left ( 1 + 2 \frac { f^{\si\ta}(r) - f^{t\ta}(r)}{f^J(r)}
 \right ) \right.  \\ & + & \left.  \left.  (\si \cdot \hat{\bf r}) (
 \hat{\bf r} \times \hat{\q} ) \left ( 1 + 6 \frac { 3
 f^{\si\ta}(r)-f^{t\ta}(r)}{f^J(r)} \right ) \right ] \right \} ~~.
\label{eq:csi_MDe}
\end{eqnarray}

 It is worth noticing that the second integral in the r.h.s. of
 Eq.(\ref{eq:csi_MDd}) gives a correction independent on the modulus
 of $\q$ and vanishing for uncorrelated wave functions as well as for
 Jastrow correlated ones.

 Fig.\ref{fig:2} compares the $f_{1,2}^\Delta$ functions obtained with
 the Argonne $v_{28}$ potential, $f_{1,2}^\Delta(A_{28})$, with those
 from the OPE potential, $f_{1,2}^\Delta(\pi)$. They differ from each
 other at short distances, and are therefore expected to provide
 similar results in a correlated model.

{\bf IV. RESULTS AND DISCUSSION}

 In this section we present and discuss the inclusive transverse
 electromagnetic response of symmetric nuclear matter at saturation
 density, $\rho_0=.16 fm^{-3}$ or $k_F=1.33 fm^{-1}$, as obtained in
 CBF theory.  We recall that the correlation operator contains
 Jastrow, spin, tensor and isospin components. The correlations are
 fixed by minimizing the ground state energy, computed in FHNC/SOC and
 using the Argonne $v_{14}$+Urbana VII three nucleon
 interaction\cite{UrbanaVII}.
 The SPR model for the two-body current operators of
 Ref.s\cite{SPR_MEC,Schiavilla92} has been used and the one-body
 current is given by the magnetization part alone, neglecting the
 convection term.  The response is computed including correlated
 $1p1h$ intermediate states and then folding $\RTPH$ with a
 parametrization of the on shell imaginary part of the CBF optical
 potential.

 We begin by studying the effect of the Jastrow, short range
 correlations and comparing with the free Fermi gas. The $1p1h$
 responses in both models are given in Fig.\ref{fig:3} at $q=300$ (a),
 $400$ (b) and $550$ (c) $MeV/c$.  In the Figures the one-body
 response (IA), the interference terms between one-body and $\JPSC$
 (OB/C), one-body and $\JPSP$ (OB/$\pi$), one-body and $\JMD$
 (OB/$\Delta$) and the quadratic $\JPSC$ (C/C) term are shown,
 together with the resulting total response (TOT).

 The shift of the Jastrow responses to higher energies respect to FG
 is due to the use of the CBF real part of the optical potential;
 moreover, as in the case of the longitudinal response\cite{R_L},
 Jastrow correlations quench the quasielastic peaks in absolute
 magnitude.  The computed quadratic term is always negligible and will
 be disregarded in the remaining of the paper. Two interference
 responses (OB/C and OB/$\Delta$) have been evaluated also for the one
 pion exchange currents.  The results at $q=400~MeV/c$ are displayed
 in Fig.\ref{fig:3}(b) as up and down triangles, respectively. They
 are almost coincident with the SPR responses, as the differences in
 the short range behaviors of the $\Delta$ currents are washed out by
 the correlation.
 
 The numerical accuracy of the FG responses has been checked against
 the analytical approach of Ref.\cite{Amaro94} finding complete
 agreement.  So, an overall reduction of the $1p1h$ QE peak due to
 MEC's is confirmed in the FG model and found even in the Jastrow
 correlated case.

 As we have already mentioned in the Introduction, this result is in
 sharp contrast with realistic estimates of the transverse response in
 light nuclei\cite{Euclid} and also with some recent RPA calculations
 in $^{12}$C\cite{RPA_MEC}, both of them pointing to a $\sim 20 - 30
 \%$ increase of the strength in the quasielastic region due to MEC
 effects.  We stress once again that this conflict is not resolved by
 the inclusion of state-independent, scalar, short range correlations.

 In Ref.\cite{R_spin} it was shown that the transverse spin responses,
 making up $\RTIA$, may be affected by the non-Jastrow correlations.
 In particular, the biggest effect was found in the isovector
 $S_{T,\si}^{\tau=1}$, as the leading correction is proportional to
 the large tensor-isospin correlation, $f^{t\tau}(r)$.  Hence, the
 correlation operator (\ref{eq:f_ij}) has been used to estimate the
 corrections to the Jastrow response due to the non scalar components,
 in the D2B/L approximation.

 Fig.\ref{fig:4}(a) reports the correlation operator results for the
 interference $1p1h$ responses at $q=400~MeV/c$. We find that, at the
 QE peak, the operatorial correlations quench the OB/C and OB/$\pi$
 responses respect to the Jastrow case. The use of the D2B in place of
 the D2B/L approximation does not change appreciably the outcome.  The
 quenching is more pronounced for the OB/$\pi$ term, where a positive
 tail is added at large energies.

 The effect is dramatic in OB/$\Delta$, as the correlation operator
 correction largely cancels the Jastrow response, yielding a positive
 net result. The origin of this cancellation is found in the
 tensor-isospin correlation contribution to the second integral in the
 r.h.s. of Eq.(\ref{eq:csi_MDd}). In fact, the OB/$\Delta$ response
 obtained by setting $f^{t\tau}=0$ in the integral is much closer to
 the Jastrow curve, as it is shown in the Figure by the $\times$
 signs.  The convergence of the cluster expansion has been checked by
 computing the integral {\it i}) in D2B approximation and {\it ii})
 adding Jastrow dressed three-body non factorizable diagrams, linear
 in the operatorial components of the correlation. The result is
 practically indistinguishable from the D2B/L response and it is not
 given in the Figure.

 A similar (and even more enhanced) behavior is found at larger
 momenta.  In Fig.\ref{fig:4}(b) we show the interference terms at
 $q=550~MeV/c$.  Now the OB/C response does not significantly differ
 from the Jastrow estimate, whereas OB/$\pi$ and OB/$\Delta$ are
 largely modified by the insertion of the operatorial correlations. In
 particular, we obtain a large, positive OB/$\Delta$ component in
 place of a large, negative Jastrow counterpart. At this momentum
 transfer and using only long range RPA correlations, the Gent group
 seems to find a more negative OB/$\pi$ response and an almost
 vanishing OB/$\Delta$ one\cite{Vander_priv}.

 The sensitivity of the OB/$\Delta$ response to the shape of
 $f^{t\tau}$ has been examined by using the correlation operator
 corresponding to the Urbana $v_{14}$+TNI model. This interaction has
 a weaker tensor component, producing a smaller tensor
 correlation. The results for OB/$\Delta$ are displayed in
 Fig.s\ref{fig:4} as full circles and point to a clear dependence on
 the details of this correlation.

 To conclude the analysis of the $1p1h$ response, we give in Figures
 \ref{fig:5} the total $\RTPH$ together with the IA estimate.  The
 total (TOT) response includes interference contributions also from
 the $\JV$ and $\JVS$ currents, whereas the curves labelled as
 TOT$_{PS}$ only contain $\JPS$ (as the Jastrow, TOT$_J$, points). MEC
 provide an enhancement of the IA response in the QE region ranging
 from $\sim 20\%$ at $q=300~MeV/c$ to $\sim 10\%$ at
 $q=550~MeV/c$. The enhancement is due to the presence of tensor- and
 isospin-dependent correlations; scalar Jastrow correlations actually
 produce MEC contributions quenching the corresponding IA values. The
 exchange of $\rho$-like vector mesons, originating the V and VS
 currents, results in an additional enhancement of $\RTPH$, which,
 however, decreases with $q$.

 Finally, we compare the transverse NM response with some of the
 available experimental data from real life nuclei. $\RT$ is obtained
 by folding the computed $\RTPH$ with the imaginary part of the
 optical potential, as described in Section II. Figures \ref{fig:6}
 compare $A\times \RT$ of NM in IA and IA+MEC with data from
 Ref.\cite{Meziani84} ($\times$), Ref.\cite{Yates93} (circles) and
 Ref.\cite{Jourdan96} (black circles) for $^{40}$Ca. The Bates data
 (Ref.\cite{Yates93}) actually refer to $q=330~MeV/c$ in
 Fig.\ref{fig:6}(a) and $q=400~MeV/c$ in Fig.\ref{fig:6}(b), as the
 Saclay data (Ref.\cite{Meziani84}) in the same Figure.

 The discrepancies between the Saclay and the Bates and Jourdan
 (Ref.\cite{Jourdan96}) results are evident and already stressed in
 previous literature. It must be noticed that the last two sets of
 data appear to be compatible among each other. This fact gives
 confidence in the Jourdan's L/T separation procedure, which employs
 world data on inclusive quasi-elastic electron scattering as obtained
 by different experiments.

 The NM responses are closer to the results of
 Ref.s\cite{Yates93,Jourdan96} than to those of Ref.\cite{Meziani84}
 and the experimental QE peak lies between the IA and the full
 responses. At $q=300~MeV/c$ IA seems to better reproduce the peak,
 whereas, when moving to higher momenta, the inclusion of MEC improves
 the agreement with the experiments.

 A similar trend is found in $^{56}$Fe, as shown in Figures
\ref{fig:7}.

{\bf V. CONCLUSIONS}

 The inclusive electromagnetic transverse response of symmetric
 nuclear matter has been evaluated within the correlated basis
 function perturbation theory. The adopted correlation operator has a
 scalar, short range component and important tensor- and
 isospin-dependent parts, its structure being similar to that of
 realistic nucleon-nucleon potentials.  Both one-body and two-body
 meson exchange currents are considered, the latter in a model which
 satisfies the continuity equation with the Argonne $v_{14}$ potential
 and contains intermediate $\Delta$-isobar excitation currents. Ground
 and $1p1h$ correlated states are included and the decay into $2p2h$
 states is implemented by folding $\RTPH$ with the imaginary part of
 the optical potential.

 Our results indicate that MEC, evaluated in a Jastrow correlated
 model, quench the IA response. In this case, the situation is
 qualitatively close to what was found by Amaro {\it et al.} in
 Ref.\cite{Amaro94} in both shell and Fermi gas models. The net
 quenching mainly originates from a strong cancellation between the
 positive contact and the negative $\Delta$ terms.

 The introduction of tensor-isospin-dependent correlations drastically
 changes this picture. The $\Delta$ contribution is largely modified,
 as it becomes positive and increasing with the momentum transfer. As
 a result, MEC produce an extra-strength (10-20$\%$) in the QE peak
 region. This is in agreement with exact GFMC calculations in light
 nuclei.

 $\rho$-like exchange currents give a small additional enhancement.
 We also found that using standard one-boson exchange currents does
 not significantly change our results.

 Two recently derived experimental responses in $^{40}$Ca have
 consistently lowered the QE peak respect to previous estimates.  The
 new data and the CBF NM responses are in reasonable agreement and the
 comparison seems to show too large MEC effects at low momenta.  The
 obvious caveat to bear in mind is that this comparison is made
 between finite nuclear systems and infinite, homogenous nuclear
 matter. CBF has been recently extended \cite{Arias96} to deal with
 ground state properties of nuclei as heavy as $^{208}$Pb, with
 Jastrow and isospin-dependent correlations.  It is conceivable that,
 in the near future, will be possible to use the theory to
 microscopically compute the finite nuclei responses, employing richer
 correlations, as those of nuclear matter. Presently, the density
 dependent NM results might be used in local density approximation for
 a closer comparison with the experiments.

 Moreover, relativistic corrections could affect the response, both
 for the IA and MEC\cite{Blunden89}, as well as the inclusion of
 explicit $\Delta$ degrees of freedom in the nuclear wave
 function. This last topic may be addressed within CBF theory,
 particularly if one wants to quantitatively study the dip and
 $\Delta$ peak regions.  In this respect, it will also be necessary to
 consider the contribution from $2p2h$ intermediate correlated states.
 Work along these lines is in progress.

{\bf ACKNOWLEDGMENTS} The author wants to thank Rocco Schiavilla for
many fruitful discussions and Juerg Jourdan for providing his data on
Ca and Fe.  The warm hospitality of TJNAF (ex CEBAF) laboratory, where
part of this work has been done, is also gratefully acknowledged.

{\bf APPENDIX}

 In this Appendix the particle-hole non diagonal matrix elements of
 the components of $\JPSP$ are given in D2B/L approximation:

\begin{eqnarray}
\nonumber {\bf \xi}^{dir}_{PS,\pi}(G_{PS,1}) & = & \delta_{{\bf
q}-{\bf p}+{\bf h}} \imath C_{PS} \frac { 6 \ta_z}{\sqrt{D(p)D(h)}} \\
& & \rho_0 \int d{\bf r} \frac {G_{PS,1}({\bf r})}{r^2} [ g_{dd}(r) +
g_{de}(r) ] e^{\imath \q \cdot {\bf r}/2} \left [ 6\imath (\si \times
\hat{\bf r}) \frac {f^{t\tau}(r)}{f^J(r)} \right ]~~,
\end{eqnarray}

\begin{eqnarray}
\nonumber {\bf \xi}^{exch}_{PS,\pi}(G_{PS,1}) & = & \delta_{{\bf
q}-{\bf p}+{\bf h}} \imath C_{PS} \frac { 6 \ta_z}{\sqrt{D(p)D(h)}} \\
& & \rho_0 \int d{\bf r} \frac {G_{PS,1}({\bf r})}{r^2} g_{cc}(r)
e^{\imath {\bf Q} \cdot {\bf r}} \left [ 2 \hat {\bf r} \left ( \frac
{5 f^{\sigma\tau}(r)- 8 f^{t\tau}(r)}{f^J(r)} \right ) - \imath (\si
\times \hat{\bf r}) \frac {f^{t\tau}(r)}{f^J(r)} \right ] ~~,
\end{eqnarray}

\begin{eqnarray}
\nonumber {\bf \xi}^{dir}_{PS,\pi}(G_{PS,2+3}) & = & \delta_{{\bf
q}-{\bf p}+{\bf h}} \imath C_{PS} \frac { 6 \ta_z}{\sqrt{D(p)D(h)}} \\
\nonumber & & \rho_0 \int d{\bf r}\left [ \frac {G_{PS,2}({\bf r}) +
G_{PS,3}({\bf r})}{r} \right ] [ g_{dd}(r) + g_{de}(r) ] e^{\imath \q
\cdot {\bf r}/2} \\ & & \left [ [ 3 (\q \cdot \hat{\bf r}) (\si \times
\hat{\bf r}) - 3 \hat{\bf r} (\si \times \hat{\bf r})\cdot \q - 2 (\si
\times \q ) ] \frac {f^{t\tau}(r)}{f^J(r)} - 2 (\si \times \q ) \frac
{f^{\si\tau}(r)}{f^J(r)} \right ] ~~,
\end{eqnarray}

\begin{eqnarray}
\nonumber {\bf \xi}^{exch}_{PS,\pi}(G_{PS,2+3}) & = & \delta_{{\bf
q}-{\bf p}+{\bf h}} \imath C_{PS} \frac { 6 \ta_z}{\sqrt{D(p)D(h)}} \\
\nonumber & & \rho_0 \int d{\bf r}\left [ \frac {G_{PS,2}({\bf r}) +
G_{PS,3}({\bf r})}{r} \right ] g_{cc}(r) e^{\imath {\bf Q} \cdot {\bf
r}} \\ & & \left [ (\si \times \q ) - 3 [ (\q \cdot \hat{\bf r}) (\si
\times \hat{\bf r}) + ( \si \cdot \hat{\bf r} ) (\q \times \hat{\bf
r}) ] \frac {f^{t\tau}(r)}{f^J(r)} \right ] ~~,
\end{eqnarray}

\begin{eqnarray}
\nonumber {\bf \xi}^{dir}_{PS,\pi}(G_{PS,4+5}) & = & \delta_{{\bf
q}-{\bf p}+{\bf h}} \imath C_{PS} \frac { 6 \ta_z}{\sqrt{D(p)D(h)}} \\
\nonumber & & \rho_0 \int d{\bf r}\left [ \frac {G_{PS,4}({\bf r}) +
G_{PS,5}({\bf r})}{r} \right ] [ g_{dd}(r) + g_{de}(r) ] e^{\imath \q
\cdot {\bf r}/2} \\ & & \left [ \hat{\bf r} (\si \times \hat{\bf
r})\cdot \q \left ( \frac {f^{t\tau}(r)- 2 f^{\si\ta}(r)}{f^J(r)}
\right ) \right ] ~~,
\end{eqnarray}

\begin{eqnarray}
\nonumber {\bf \xi}^{exch}_{PS,\pi}(G_{PS,4+5}) & = & - \delta_{{\bf
q}-{\bf p}+{\bf h}} \imath C_{PS} \frac { 6 \ta_z}{\sqrt{D(p)D(h)}} \\
& & \rho_0 \int d{\bf r}\left [ \frac {G_{PS,4}({\bf r}) +
G_{PS,5}({\bf r})}{r} \right ] g_{cc}(r) e^{\imath {\bf Q} \cdot {\bf
r}} \left [ \hat{\bf r} (\si \times \hat{\bf r})\cdot \q \right ] ~~,
\end{eqnarray}

\begin{equation}
{\bf \xi}^{dir}_{PS,\pi}(G_{PS,6}) = 0 ~~,
\end{equation}

\begin{eqnarray}
\nonumber {\bf \xi}^{exch}_{PS,\pi}(G_{PS,6}) & = & - \delta_{{\bf
q}-{\bf p}+{\bf h}} \imath C_{PS} \frac { 6 \ta_z}{\sqrt{D(p)D(h)}} \\
& & \rho_0 \int d{\bf r} G_{PS,6}({\bf r}) g_{cc}(r) e^{\imath {\bf Q}
\cdot {\bf r}} \left [ 3 \imath \hat{\bf r}(\q \cdot \hat{\bf r})(\si
\times \hat{\bf r})\cdot \q \frac {f^{t\tau}(r)}{f^J(r)} \right ] ~~,
\end{eqnarray}

\begin{equation}
{\bf \xi}^{dir}_{PS,\pi}(G_{PS,7}) = {\bf
\xi}^{exch}_{PS,\pi}(G_{PS,7}) = 0 ~~,
\end{equation}

 where ${\bf Q}=(\p + \h)/2$.

 \begin{figure} \caption{$g_{PS}$ functions for the Argonne $v_{14}$,
 $g_{PS}(A_{14})$, and the $\pi$ exchange, $g_{PS}(\pi)$,
 potentials. $g_{cc}(FHNC)$ and $g_{cc}(FG)$ are the correlated and
 uncorrelated exchange functions, respectively (see text).}
 \label{fig:1} \end{figure}
 
 \begin{figure} \caption{$f_{1,2}^\Delta$ functions for the Argonne
 $v_{28}$, $f_{1,2}^\Delta(A_{28})$, and the $\pi$ exchange,
 $f_{1,2}^\Delta(\pi)$, potentials.}  \label{fig:2} \end{figure}
 
 \begin{figure} \caption{$1p1h$ transverse response at $q=300$ (a),
 $400$ (b) and $ 550$ (c) $MeV/c$ for the FG and Jastrow models. The
 Figure shows the partial (one-body, interference and one quadratic)
 and the total responses. The interference terms for the contact
 (triangles up) and $\Delta$ (triangles down) currents in the one pion
 exchange model at $q=400 MeV/c$ are given.}  \label{fig:3}
 \end{figure}

\begin{figure}
\caption{$1p1h$ interference responses at $q=400$ (a) and $550$ (b)
 $MeV/c$ in the correlation operator model and comparison with the
 Jastrow model.  See text.}
\label{fig:4}
\end{figure}

\begin{figure}
\caption{Total $1p1h$ responses at $q=300$ (a), $400$ (b) and $550$
 (c) $MeV/c$ in the correlation operator model and comparison with the
 Jastrow model and the IA responses.  See text.}
\label{fig:5}
\end{figure}

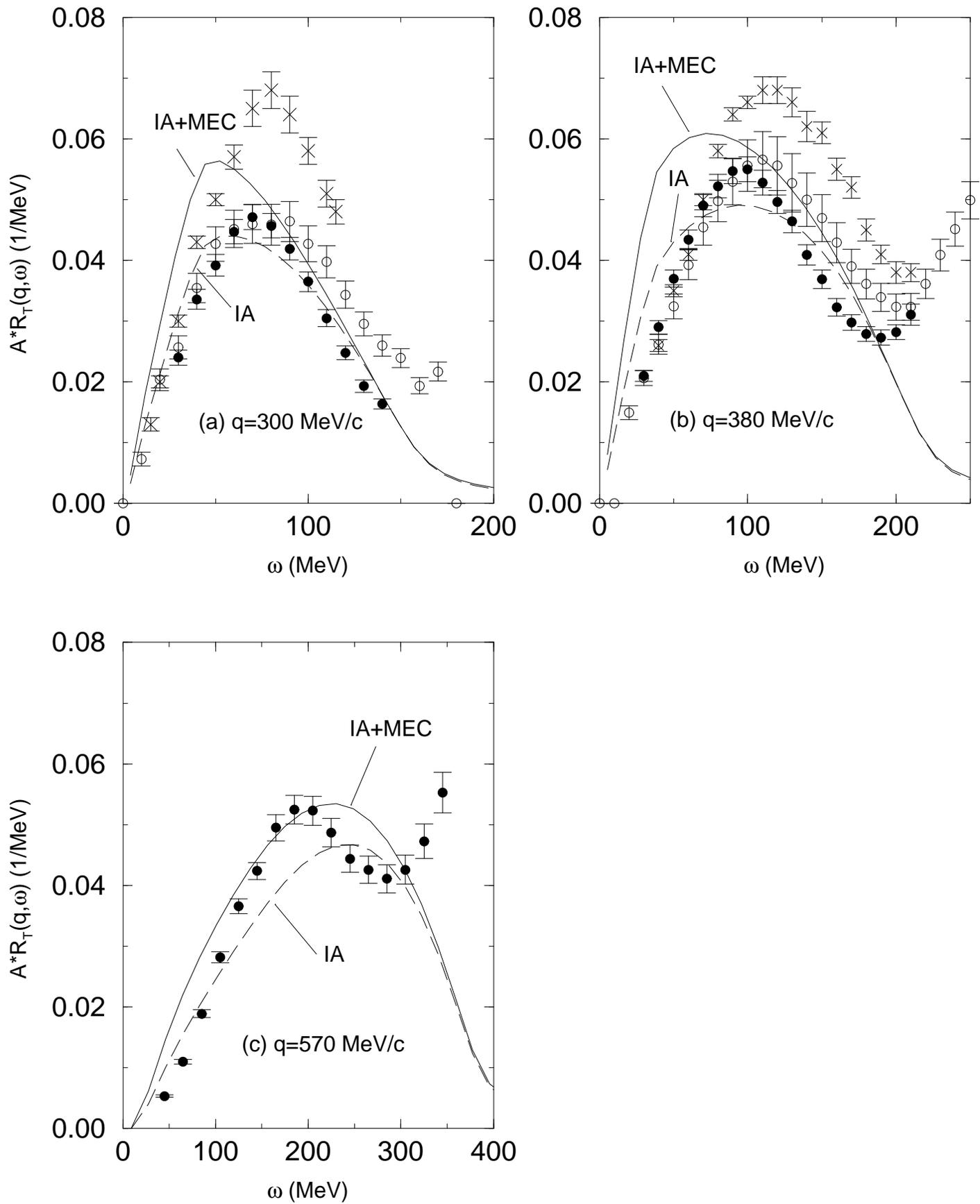
\begin{figure}
\caption{ Transverse responses at $q=300$ (a), $380$ (b) and $570$ (c)
 $MeV/c$ for $^{40}$Ca and nuclear matter.  See text.}
\label{fig:6}
\end{figure}

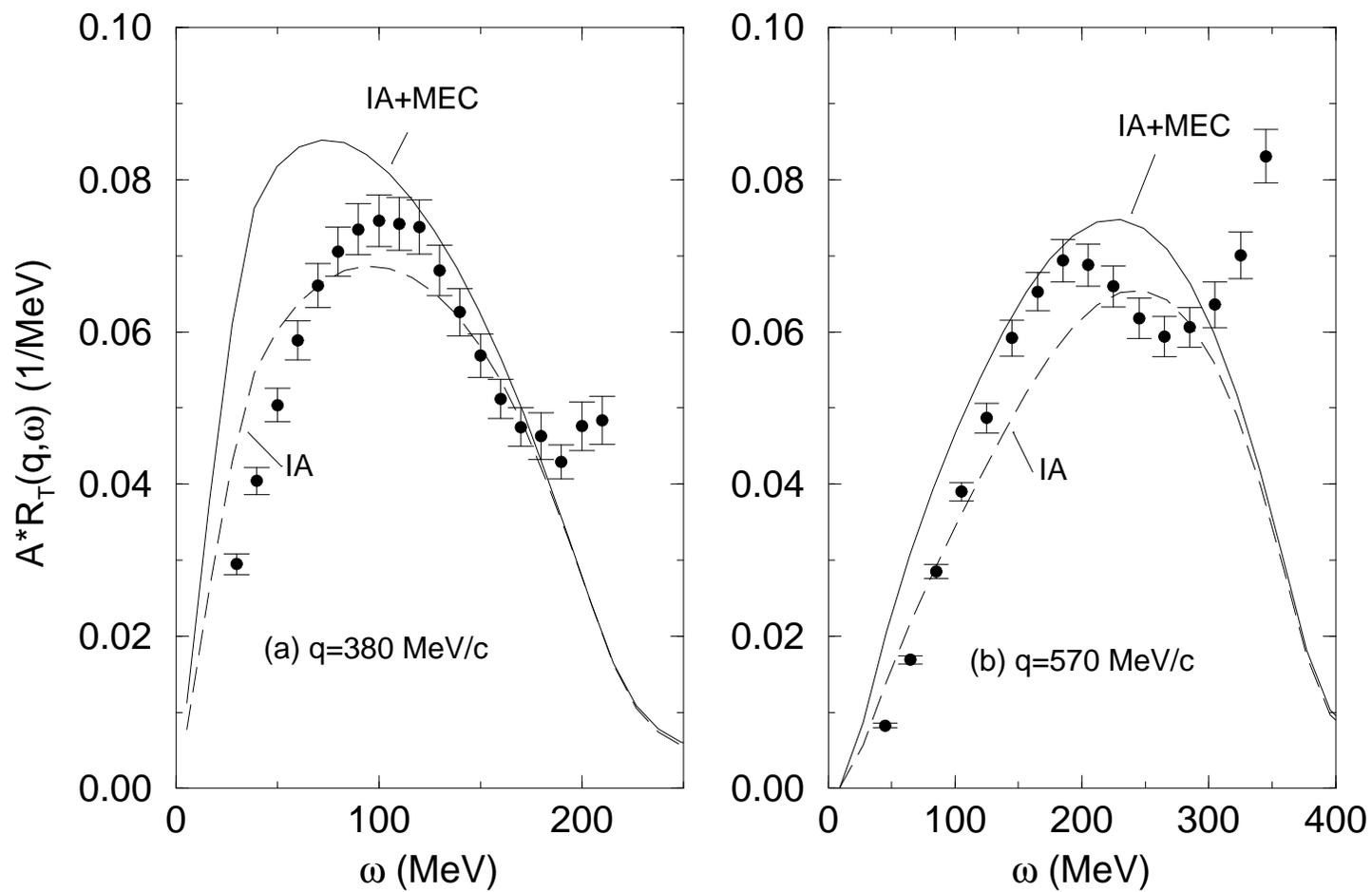
\begin{figure}
\caption{ Transverse responses at $q=380$ (a) and $570$ (b) $MeV/c$
 for $^{56}$Fe and nuclear matter.  See text.}
\label{fig:7}
\end{figure}

\end{document}